\documentstyle[floats,aps]{revtex}
\begin{document}
\twocolumn[\hsize\textwidth\columnwidth\hsize\csname  
@twocolumnfalse\endcsname

\title{Conductance quantization and electron resonances in
sharp tips and atomic-size contacts}

\author{A. Levy Yeyati, A. Mart\'{\i}n-Rodero, F. Flores}

\address{
Departamento de F\'\i sica  Te\'orica de la Materia Condensada C-V.\\
Universidad Aut\'onoma de Madrid, E-28049 Madrid, Spain.}

\maketitle

\begin{abstract}
The electronic and transport properties of atomic-size contacts are
analyzed theoretically using a self-consistent tight-binding
model. Our results show that, for $s$-like metals, a sufficiently
narrow contact exhibits well defined resonant states at the
Fermi energy, spatially localized in the neck region. These states are
robust with respect to disorder and provide a simple explanation for the
observed tendency to conductance quantization.
It is also shown that these properties disappear for
a sufficiently large contact area. The possible relevance of the
resonant states in scanning tunneling spectroscopy using sharp tips is
briefly discussed. 
\end{abstract}

\vskip2pc]

\narrowtext

The electronic and transport properties of metallic point contacts 
have received a renewed attention with the advent of controllable
atomic-size contacts that can be produced using the scanning
tunneling microscope \cite{STM} and the more recently developed 
mechanically controllable break junctions \cite{MCBJ}. 

By means of these techniques, conductance steps 
in both the jump-to-contact and jump-to-tunnel processes are generally
observed in the experiments \cite{STM,MCBJ,steps}.
The steps have been correlated with abrupt atomic rearrangements 
taking place in the neck region \cite{STM}.
An important issue, still not completely understood \cite{discussion}, 
is whether these steps should correspond to quanta of the conductance. 
What seems clear from the experimental data is that,
for metals having basically an $s$-character ($Na$, $Au$, $Ag$, $Cu$, etc),
the value of the first conductance jumps 
tends to be very close to a multiple of the quantum of conductance $2e^2/h$,
while the higher conductance steps exhibit a less reproducible behavior
\cite{steps}.
On the other hand,
transition metals like $Ni$ and $Pt$ exhibit a broad distribution
of the first conductance step, which can be larger than $2e^2/h$ by a factor 
between 1 and 3. 
This effect has been attributed to the role played by the $d$-electrons
of these materials that open new channels to the electronic conduction
\cite{Sirvent}.

Theoretically, different authors have analyzed the mechanical properties
of these atomic-size contacts using molecular dynamics
simulations \cite{MD} as well as their transport properties by means 
of tight-binding \cite{TB} and free-electron calculations \cite{FE}. 
The molecular dynamics simulations show that 
a one-atom point contact is a stable configuration \cite{MD} and
several theoretical models 
have predicted that the maximum value of the conductance is $2e^2/h$ for a
single atom contact with one orbital per site 
\cite{Lang,Ferrer,Ciraci,Sutton}. Nevertheless, it is not yet clear
why the first step quantization should be observed in such a 
reproducible way for simple metal contacts.
Free-electron calculations have tried to explain this fact by using an
adjustable contact cross section, $\pi r_0^2$, taking into account the size 
of the contact atom \cite{FE}. 
However, the results of this model seem to depend critically on 
the ratio $\lambda_F/r_0$.
On the other hand, conductance quantization seems to be even much harder
to obtain using a tight-binding description of realistic models for the
contact atomic structure \cite{TB}.

In this work we address 
this problem by focusing our attention
on the electronic properties of very sharp tips 
and narrow atomic-size contacts.  
We concentrate our discussion on metals having 
essentially an $s$-like band character
which, as commented above, are the ones displaying conductance
quantization in a more reproducible way.
We shall show that for sufficiently sharp tips there appear
narrow electron resonances which are spatially localized around the apex atom.
As discussed below, 
these states could certainly play an important role in scanning
tunneling spectroscopy. 
When forming an atomic contact, and due to charge neutrality
conditions, the
Fermi level is pinned at one of these resonances, explaining in a
natural way conductance quantization at the first plateaux.
Furthermore, we shall show
that these results are robust with respect to the 
inclusion of disorder.  

In order to describe the electronic and transport properties of an
atomic-size tip or contact, we shall use an atomic
orbital basis leading to a tight-binding Hamiltonian with the form

\begin{eqnarray}
\hat{H} & = & \sum_{i\alpha,\sigma} \epsilon_{i\alpha} 
c^{\dagger}_{i\alpha,\sigma} c_{i\alpha,\sigma}
+ \sum_{i\alpha \neq j\beta, \sigma} t_{i\alpha,j\beta} 
c^{\dagger}_{i\alpha,\sigma} c_{j\beta,\sigma},
\end{eqnarray}

\noindent
where indexes $i,j$ run over the atomic sites and $\alpha,\beta$
denote the different atomic orbitals at each site. 
For representing simple metals,
we shall first consider the case of a single $s$ orbital per site and 
then discuss qualitatively the effect of including $p$ orbitals.
The hopping elements
$t_{i\alpha,j\beta}$ are assumed to connect first-neighboring sites,
and can be taken from fittings to ab-initio electronic band calculations
\cite{Papa}. When considering a single orbital per site the relevant
parameter is the first-neighbors hopping element $t$, which is of the
order of $1 eV$ for typical metals.
The diagonal elements $\epsilon_{i\alpha}$ will be modified self-consistently
to achieve local charge neutrality, which is a reasonable assumption for
a metallic system \cite{Pernas}. 

The first step in this analysis is to show that sharp resonant states
can exist in a tip geometry (i.e., a semi-infinite arrangement of
atoms like the ones depicted in Fig. 1) provided it is sufficiently sharp.
Physical intuition suggests that metallic
atomic tips and contacts would most likely have a close packed structure.
This is supported by computer simulations \cite{MD} at least for sufficiently
large temperature, when the tendency to minimize the contact energy can
be developed. 
We thus first consider tip geometries which can be grown from an apex atom
along different crystallographic orientations on an fcc lattice, adding
the first neighboring sites when going from one atomic layer to the
next. 
In this way, when the (111) direction is chosen, one obtains a very
sharp tip with an opening angle of $\sim \pi/4$; we shall also consider
the (100) direction that leads to a tip with a larger opening angle
of $\sim \pi/2$.

We have studied the electronic structure of these tips by first
considering small clusters around the apex atom with an increasing 
number of layers $n$. The atoms on the last layer are connected
to a Bethe lattice structure that simulates the bulk DOS. The inset of Fig. 1
shows the LDOS at the tip apex atom as a function of $n$ for a (111)
tip. 
These results show that, as $n$ increases, the spectral density for $E
\ge 0$ tends to concentrate into two sharp resonances at $E \sim 0.4 t$
and $\sim 1.5 t$. The LDOS below $E = 0$ is,
however, much more sensitive to the number of layers in the cluster
and converges slowly to a continuous spectrum, as corresponds to
extended states. This part of the spectrum is thus not completely well
described by a small cluster calculation. 

A more efficient algorithm for calculating the LDOS at a given atom 
is provided by the recursion method \cite{Haydock}.
Using this method one obtains an expansion of the local Green functions
as a continued fraction

\begin{eqnarray}
G_{j\alpha,j\alpha}(\omega) & = & \frac{1}{\omega - a_0 - b_0^2
g_1(\omega)}, \nonumber \\
g_n(\omega) & = & \frac{1}{\omega - a_n - b_n^2 g_{n+1}(\omega)}
\end{eqnarray}

\noindent
where the coefficients $a_n,b_n$ are obtained by generating a new
basis recursively \cite{Haydock} taking a generic 
local orbital $|j\alpha>$ as the
initial state. Notice that the coefficients $a_n,b_n$ define an
effective 1D tight-binding Hamiltonian  
with diagonal levels $a_n$ and hopping elements $b_n$.
When starting from the apex site ($j=0$), 
the $n$ site in this effective semi-infinite chain corresponds to a 
state of the tip having a finite weight on the first $n$ layers. 

The LDOS at the tip apex site obtained for the (100) and (111) grown
tips are shown in Fig 1. 
As can be observed, the less sharp tip exhibits a smooth LDOS
resembling the fcc bulk DOS. On the contrary, for the sharper (111) case
the LDOS is qualitatively different, exhibiting 
narrow resonances in agreement with those found in the finite cluster
calculation. 

The appearance of these resonances can be understood in the following
way: while in a blunt tip the local environment around the apex atom
is similar to that of a flat surface tending quickly to that of the
bulk, in the sharp (111) structure it converges more slowly and thus the
apex region, which is much more weakly coupled to the bulk, can exhibit
resonant states. This is reflected
in the different evolution of the $a_n,b_n$ coefficients towards their
bulk values $(a_{\infty} = -4t, b_{\infty} = 4t)$, as shown in Fig. 2.
Furthermore, the coefficients of the (111) case exhibit significant
fluctuations which decrease rather slowly with increasing $n$. 
The effective 1D Hamiltonian for this case thus describes the motion 
of an electron in a fluctuating potential, which displays a region 
of localized states.
In fact, the wave functions of the effective 1D
problem that correspond to resonant states are strongly
localized around the tip apex, with a typical localization length 
corresponding to a few atomic layers.


Broadening of the resonant levels is mainly introduced by topological
disorder, which has been simulated by allowing the 
hopping elements to fluctuate randomly within a certain range $t \pm
\Delta t$. The broadening can be 
understood by noticing that the effect of disorder
is to reduce the fluctuations in the effective 1D
parameters $a_n$ and $b_n$ due to 
a loss of coherence between the different interfering paths from the
tip apex to a given layer. 
This is illustrated in the inset of Fig. 2.
We have found, however, that this effect is never very large and
the resonances remain well defined (even for $\Delta t \sim t$, 
which is much larger than the expected fluctuations on $t$,
the peak labeled as B in Fig. 1 acquires only a small 
broadening $\sim 0.1 t$).

The relevance of these resonant states becomes clear after 
realizing that, for a (111) tip, the Fermi level is located 
above the smooth part of the apex atom LDOS. 
This is so because, the total charge per spin on the smooth LDOS
is less than $0.5$ electrons and $E_F$ turns out to be located between 
the two resonances A and B, as indicated in Fig. 1.
Accordingly, these resonances should be observed in Scanning Tunneling
Spectroscopy when using sufficiently sharp tips. 
Notice that typical tips are, however, made of transition metals like W
for which the simplified one orbital per site model is not sufficient.
We have performed specific calculations for W tips including $d$
orbitals and found that they should also exhibit resonant states in
certain conditions. 
This has been confirmed by recent experiments \cite{private} 
and might clarify an
existing controversy regarding the electronic structure of W-tips
\cite{Binh}. 
 
It seems then plausible that the same kind of resonant states could
be present in a one-atom neck geometry, which can be viewed as two tips
connected by a central common atom. This model geometry can either
represent the atomic contacts that are formed with the break-junction
technique or by using the STM techniques after repeated cycles of
plastic deformation \cite{Untiedt}.
The self-consistent LDOS at the central atom and on a neighboring layer 
of a (111) neck
are shown in Fig. 3. As expected, they exhibit resonant states, which
are, however, shifted from their position in the tip case. The LDOS at
the central atom is dominated by a clear resonance at $E \sim 0.35t$, 
whereas the resonance at $E_F$ is mainly localized on the first 
neighboring
layers.  The spatial variation of
this resonant state is depicted in the lower inset of Fig. 3.
The broadening of these states due to topological disorder is twice as
large as the one found for the tip; thus the maximum broadening is
around $0.2 t$ and the typical line-width is $\sim 0.1 t$, i.e. around
$0.1 eV$ for simple metals. 

The existence of a half-filled resonant state for the (111) neck
geometry provides a strong mechanism for explaining the almost
perfect quantization of the conductance at the first plateau. 
In fact, the conductance of
this single atom contact is given by \cite{dotpaper}

\begin{equation}
g = \frac{2e^2}{h} \left\{ 4 Im{\Sigma_R(E_F)} Im{\Sigma_L(E_F)} 
|G(E_F)|^2 \right\}
\end{equation}

\noindent
which behaves as the usual resonant tunneling transmission formula
around a resonant state \cite{dotpaper}. 
We use the zero temperature conductance, as the line-width of
the resonant state at $E_F$, although small compared to $t$, 
can be expected to be much
larger than $k_B T$, even at room temperature.
$\Sigma_R(\omega)$ and $\Sigma_L(\omega)$ are 
the self-energies of the right
and left sides of the contact projected on the central atom; $G(\omega)$
is the local Green function on this atom, given by $1/[\omega -
\epsilon_{0} - \Sigma_L(\omega) - \Sigma_R(\omega)]$.
As the Fermi energy is located at the center
of a sharp resonance, Eq. (3) takes the form
$g = (8e^2/h) x/(1 + x)^2$, 
where $x = Im[\Sigma_R(E_F)]/Im[\Sigma_L(E_F)]$. For the perfect (111)
atomic contact $x = 1$, and $g = 2e^2/h$. Fluctuations in the atomic
positions introduce variations in the values of
$Im[\Sigma_{R,L}(E_F)]$, but, as far as the resonance is well defined,
charge neutrality keeps it centered around $E_F$. 
One should notice, however, that around $x = 1$,
$g$ changes very slowly with $x$; thus, for a deviation as large
as $x = 2$, $g \simeq 1.8 e^2/h$, i.e. $90 \%$ of the quantum unit.  
Sharp resonances thus provide a strong mechanism for the conductance
quantization on the first plateau \cite{comment}. 
In the opposite case of a smooth LDOS
(like in the (100) case) perfect transmission and maximum conductance
only occur at certain energies which {\it do not}, in general,
correspond to $E_F$. 

The previous analysis has been restricted to one-atom contacts.
It is also worth considering the effect of increasing the contact area.
For instance, three atom contacts can be obtained from the initial (111)
structure by removing the central atom, thus defining the sequence
$...6-3-3-6...$, where the numbers give the atoms per layer.
This geometry also exhibits sharp resonances in the neck LDOS, allowing
us to predict, by a simple electron counting argument, 
a conductance of $2 \times (2e^2/h)$ ($E_F$ lies in the middle of
a two-fold degenerate level). A different three-atom contact
configuration like $...6-3-6...$ would correspond to a conductance of $3
\times (2e^2/h)$.  
Can we expect to obtain increasing conductance quanta,
$n \times (2e^2/h)$, when the neck contact area is progressively increased?. 
The arguments given in this work suggest that the limit to a well defined
conductance quantization should appear for a contact size for which the
resonant states at $E_F$ begin to disappear. 
In this respect, we have analyzed the LDOS of contacts with an
increasing number of atoms in the neck section and found that for
sections of the order or larger than six atoms the resonances become too
broad to be resolved. 

So far we have neglected the effect of $p$ orbitals. We have also
performed calculations including them but still assuming to have 
a single electron per site. Alkali atoms are typical cases of this situation
but also $Au$, $Ag$ and $Cu$ can be well represented by this model. 
Our results show that there is still a narrow resonance at the Fermi
level having mainly an $s$-character, giving further support
to the main argument of this work. 
For metals like $Al$, having more conduction electrons per atom
and a larger $p$-character at $E_F$, the
resonances become very much broadened. As a consequence we would not
expect the conductance quantization of the first plateau to be as robust in
these metals as in the previous cases, in agreement with the 
experimental evidence \cite{steps}.   

In conclusion we have shown that, for metals having mainly an $s$-character
and for sufficiently narrow tips, resonant
electron states appear at and around $E_F$, spatially localized at
the tip apex. The inclusion of the discrete atomic structure 
is {\it essential} for obtaining these states, which would not be
present in a jellium constriction model. 
When a single atom contact is formed, $E_F$ is pinned by
one of those resonances and a quantum unit conductance is obtained, a
result that is shown to be robust with respect to the inclusion of
disorder. By increasing the contact area, the conductance is expected to
increase by multiples of the quantum unit as far as those resonant
states still exist at $E_F$. A limit to this increase is set by the
disappearance of the resonances due to the increase of the contact size. 
Our calculations
show that for a contact area of six atoms the resonances become too
broad and explain why in most experiments the conductance quantization
$n \times (2e^2/h)$ is only seen in a reproducible way for small 
values of $n$.  

\acknowledgements
The authors would like to thank M. B\"uttiker for his remarks on this
manuscript.
Support by Spanish CICYT (contracts No. PB93-0260 and PB92-0168C) is
acknowledged.

\begin{figure}[h]
\caption{LDOS at the apex atom for tips grown in the (111) (full line)
and in the (100) (dotted line) directions. The upper insets give a top
view of the two tip geometries. In the lower inset the apex atom LDOS
for small (111) clusters are shown for $n= 6$, 8, 10 and 12 layers. 
The two resonant states are labeled by $A$ and $B$.}
\end{figure}

\begin{figure}[h]
\caption{Coefficients $a_n$ and $b_n$ in the continued fraction
expansion of the tip apex Green function for the (111) (full
line) and (100) (dotted line) geometries. In the inset the behavior of
$a_n$ with increasing amplitude of random fluctuations in the hopping
parameter, $\Delta t$, is shown for the (111) case.} 
\end{figure}

\begin{figure}[h]
\caption{LDOS at the central (full line) and at the first neighboring
layer (dotted line) for the (111) neck geometry. The broken line gives
the contact conductance. The upper inset gives a side view of the (111)
neck geometry. In the lower inset the effective 1D problem wave-function
corresponding to the resonant state at $E_F$ is shown.}
\end{figure} 

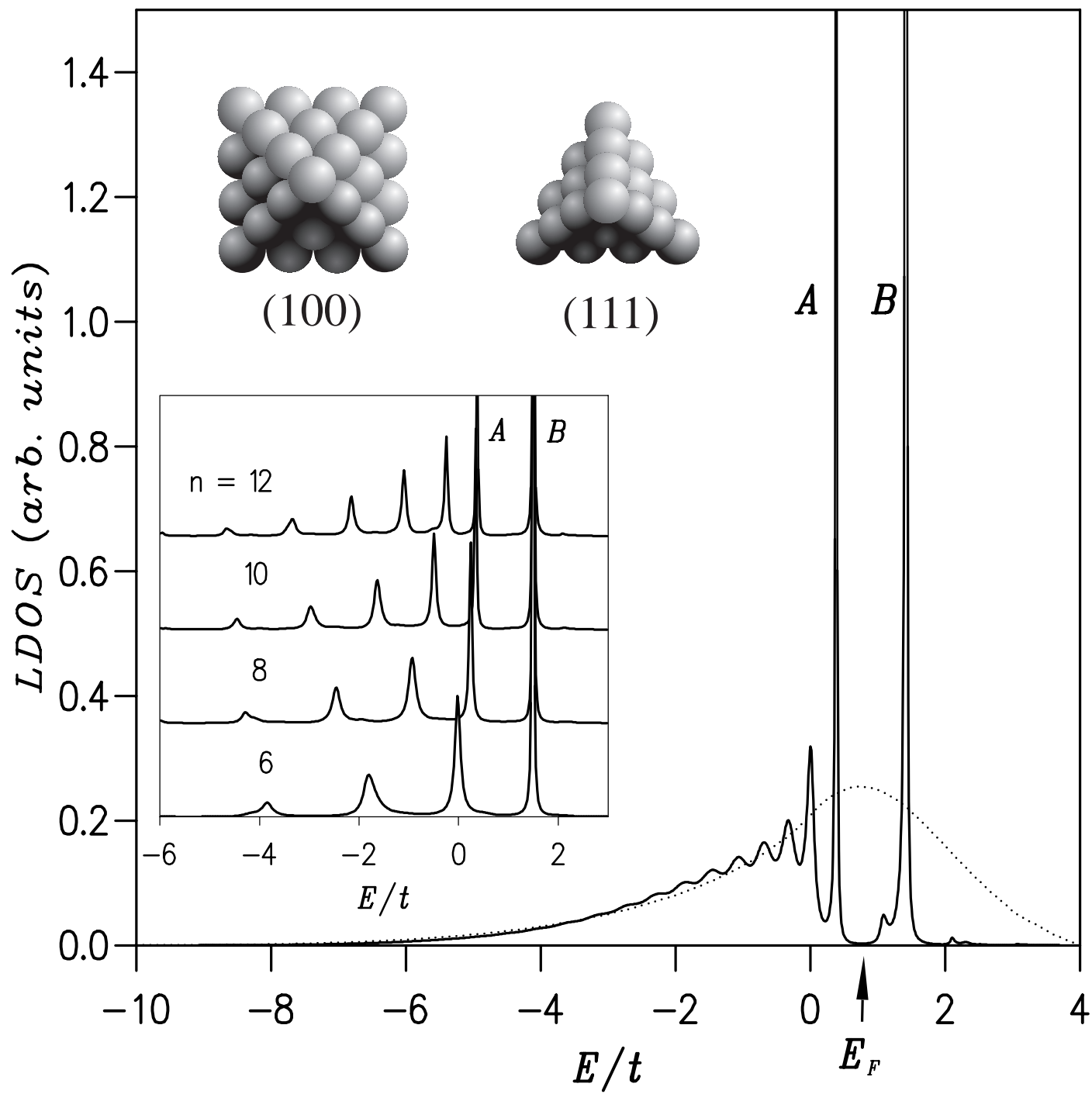
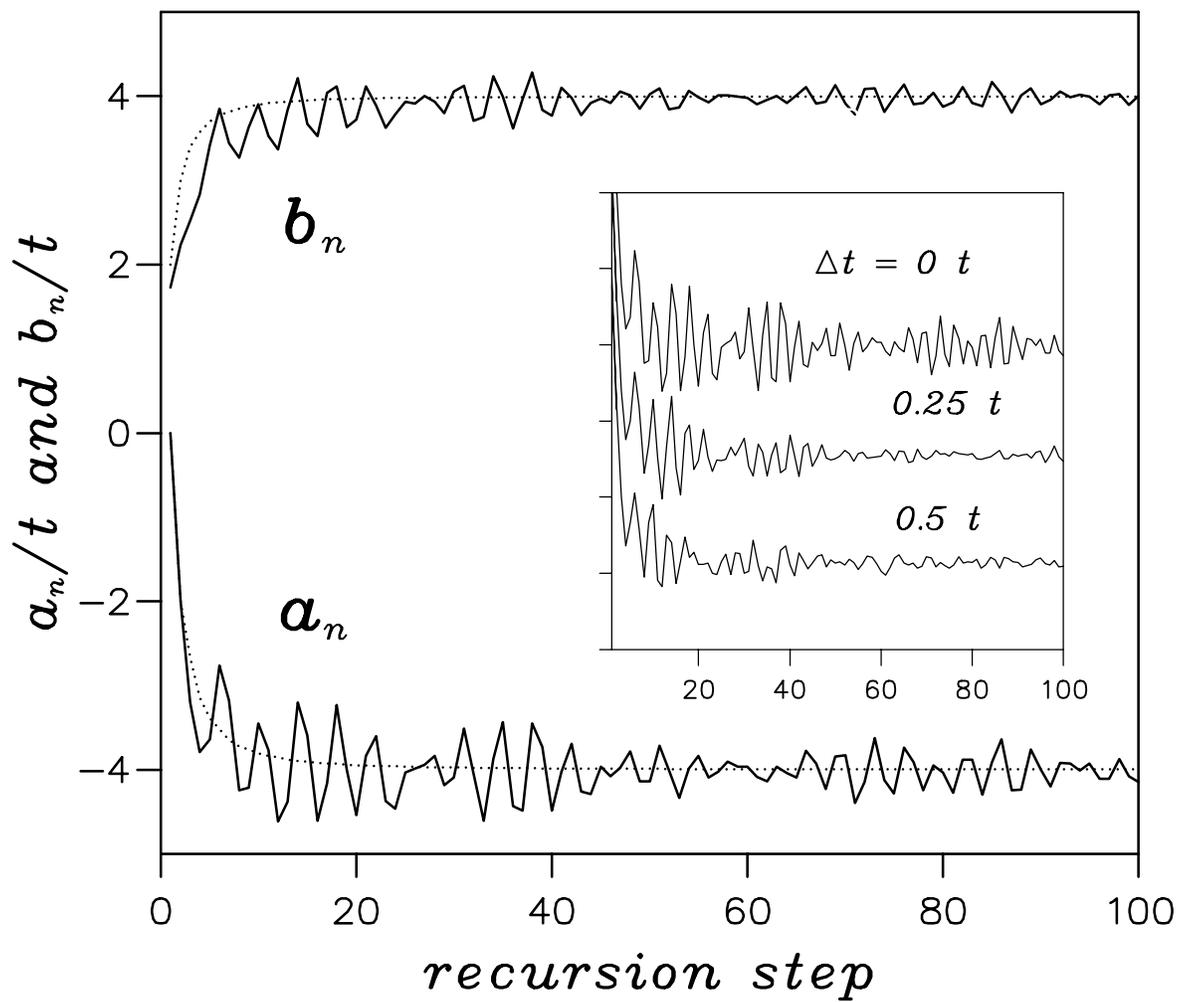
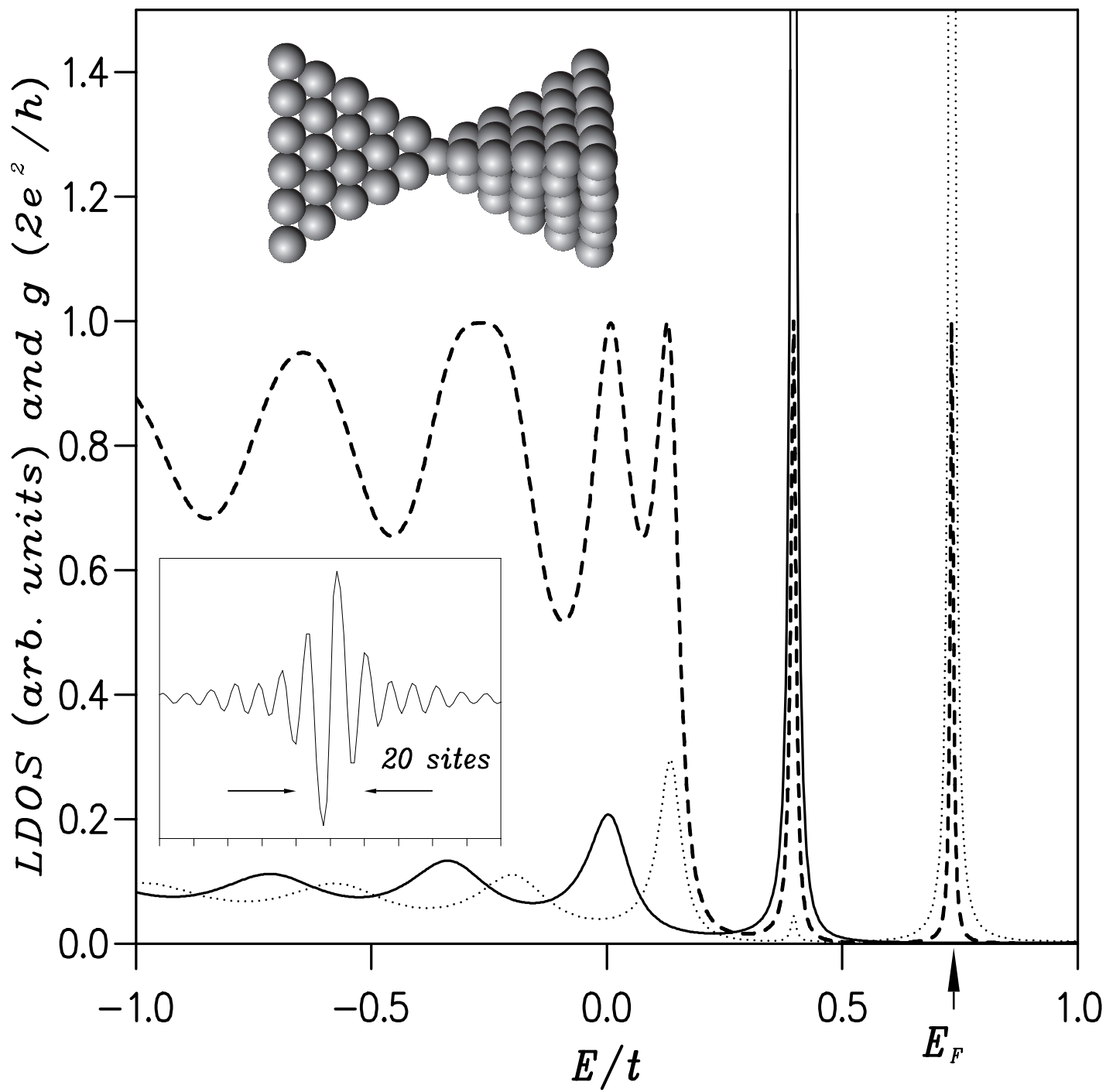
\end{document}